\documentclass[runningheads]{llncs}
\usepackage[T1]{fontenc}
\usepackage{fancyvrb,graphicx}
\usepackage{xcolor}

\newcommand{\SAD}{\ensuremath{\mathtt{SAD}}}
\newcommand{\dict}{\ensuremath{\mathtt{dict}}}
\newcommand{\dSize}{\ensuremath{\mathtt{dSize}}}
\newcommand{\valid}{\ensuremath{\mathtt{valid}}}
\newcommand{\unique}{\ensuremath{\mathtt{unique}}}
\newcommand{\occ}{\ensuremath{\mathtt{occ}}}
\newcommand{\BWT}{\ensuremath{\mathtt{BWT}}}
\newcommand{\oldBWT}{\ensuremath{\mathtt{oldBWT}}}
\newcommand{\DS}{\ensuremath{\mathtt{DS}}}
\newcommand{\buffer}{\ensuremath{\mathtt{buffer}}}

\begin{document}

\title{Prefix-free parsing for merging big BWTs}

\titlerunning{PFP for merging big BWTs}

\author{Diego D\'iaz-Dom\'inguez \inst{1}\orcidID{0000-0002-9071-0254} \and \\
Travis Gagie \inst{2}\orcidID{0000-0003-3689-327X} \and \\
Veronica Guerrini \inst{3}\orcidID{0000-0001-8888-9243} \and \\
Ben Langmead \inst{4}\orcidID{0000-0003-2437-1976} \and \\
Zsuzsanna Lipt\'ak \inst{5}\orcidID{0000-0002-3233-0691} \and \\
Giovanni Manzini \inst{3}\orcidID{0000-0002-5047-0196} \\
Francesco Masillo \inst{6}\orcidID{0000-0002-2078-6835} \and \\
Vikram Shivakumar \inst{4}\orcidID{0000-0002-1424-3711}}

\authorrunning{D. D\'iaz-Dom\'inguez et al.}

\institute{University of Helsinki, Finland \and
Dalhousie University, Canada \and
University of Pisa, Italy \and
Johns Hopkins University, USA \and
University of Verona, Italy \and
Dortmund Technical University, Germany}

\maketitle

\begin{abstract}
When building Burrows-Wheeler Transforms (BWTs) of truly huge datasets, prefix-free parsing (PFP) can use an unreasonable amount of memory.  In this paper we show how if a dataset can be broken down into small datasets that are not very similar to each other --- such as collections of many copies of genomes of each of several species, or collections of many copies of each of the human chromosomes --- then we can drastically reduce PFP's memory footprint by building the BWTs of the small datasets and then merging them into the BWT of the whole dataset.

\end{abstract}

\section{Introduction}
\label{sec:introduction}

The discovery of the r-index~\cite{GNP18,GNP20} was an exciting time in bioinformatics but, as Paolo Ferragina~\cite{FGM12} likes to say, to use an index one must first {\em build} it!  There were at that point no algorithms for building Burrows-Wheeler Transforms (BWTs) that could handle dozens of human genomes in reasonable time and memory but, inspired by their desire to use the r-index, Boucher et al.~\cite{BGKM18,BGKLMM19} quickly proposed prefix-free parsing (PFP).  Although lacking good worst-case guarantees, PFP is fast and fairly small in practice, and easy to implement: on Feb.\ 21st, 2018 Gagie sent his final design~\cite{GM18} to Manzini, who had a working implementation a week later.  These features have made PFP popular, with several groups (for example,~\cite{ARGBL23,ARKSGBL2,BCLRS21,BCGHMNR21,FOGB24,GIMNST19,GB22,HRB23,KVOB24,LLM25,OBO25,ROLGB22}) using and extending it, mostly without Manzini's help.  It is even covered in the new edition of M\"akinen et al.'s~\cite{MBCT23} textbook!

Sadly, PFP seems to have reached the limit of its usefulness now that we have truly huge pangenomic datasets to index: it ran out of memory when Li~\cite{Li24} tried to build the BWT of AllTheBacteria~\cite{HLAHSLI24}, and when the fourth and eighth authors of this paper tried to build the BWT of the latest HPRC reference of 231 human genomes (``Assemblies Release 2'' at \url{https://humanpangenome.org/data}).  These datasets are not repetitive enough for recursive PFP~\cite{FOGB24} to work well, so it seems we must turn to newer and more sophisticated algorithms, such as D\'iaz-Dom\'inguez and Navarro's~\cite{DN23} grlBWT, Li's~\cite{Li24} ropebwt3, Masillo's~\cite{Mas23} CMS-BWT or Olbrich's~\cite{Olb25} algorithm.  In this paper, however, we show there is life in PFP yet!

Specifically, we show if a dataset can be broken down into small datasets that are not very similar to each other --- such as collections of many copies of genomes of each of several species, or collections of many copies of each of the human chromosomes --- then we can drastically reduce PFP's memory footprint by building the BWTs of the small datasets and then merging them into the BWT of the whole dataset.  In Section~\ref{sec:PFP} we briefly review how PFP works, in Section~\ref{sec:merging} we describe how it can be used for merging BWTs, and in Section~\ref{sec:experiments} we demonstrate experimentally that our idea is practical.

\section{PFP}
\label{sec:PFP}

PFP is based on a modification of rsync~\cite{TM96}: we run a sliding window of a given length $w$ over the dataset and insert a phrase break whenever the Karp-Rabin hash of the contents of the window are 0 modulo a given value $p$ (which need not be prime).  Unlike rsync, however, we include the contents of the window that triggered the break --- called a {\em trigger string} --- in both the preceding and next phrases.  If we treat the string as circular then
\begin{itemize}
\item consecutive phrases overlap by exactly $w$ characters;
\item every character in the dataset is contained in exactly one phrase in which it is not among the last $w$ characters;
\item every phrase starts with a trigger string, ends with a trigger string and contains no other trigger string;
\item no proper phrase suffix of length at least $w$ --- called a {\em valid} phrase suffix --- is a proper prefix of any other proper phrase suffix (otherwise there would be a trigger string in the middle of a phrase).
\end{itemize}

It follows that if we know the lexicographic order of the dataset's suffixes starting at phrase boundaries --- and we can find that by building the BWT of the parse --- then we can easily determine the order in the BWT of any two characters in the dataset.  To do this, we consider the unique phrases in which those characters appear but not among the last $w$ characters, and compare the valid phrase suffixes immediately following the characters.  If those phrase suffixes differ, then the characters' order in the BWT is the same as the lexicographic order of the phrase suffixes; otherwise, it is the same as the lexicographic order of the dataset's suffixes starting at the ends of the phrases.

To compare valid phrase suffixes quickly, we can build the suffix array $\SAD$ of the concatenation $\dict$ of the phrases in the dictionary.  In fact, if
\begin{itemize}
\item $\dSize$ is the number of characters in $\dict$,
\item $\valid(\SAD[i])$ indicates whether the phrase suffix starting at $\dict[\SAD[i]]$ is valid,
\item $\unique(\SAD[i])$ indicates whether all occurrences in $\dict$ of the phrase suffix starting at $\dict[\SAD[i]]$ are preceded by copies of the same character,
\item $\occ(\SAD[i])$ is the frequency in the parse of the phrase containing $\dict[\SAD[i]]$,
\item $\BWT$ is a string with the same length as the dataset,
\end{itemize}
then the code fragment shown in Figure~\ref{fig:build_code} fills in the characters in the BWT that precede in the dataset valid phrase suffixes always preceded by the same distinct character.  It considers only the dictionary and not the parse, usually fills in most of the BWT by itself --- and it uses some ideas that will be useful to us in this paper.

\begin{figure}[t]
\begin{center}
\begin{BVerbatim}
int pos = 0;
for (int i = 0; i < dSize; i++) {
    if (valid(SAD[i])) {
        if (unique(SAD[i])) {
            memset(&BWT[pos], dict[SAD[i] - 1], occ(SAD[i]));
        }
        pos += occ(SAD[i]);
}   }
\end{BVerbatim}
\caption{A code fragment for filling in the characters in the BWT that precede in the dataset valid phrase suffixes always preceded by the same distinct character.  By itself, this usually fills in most of the BWT.}
\label{fig:build_code}
\end{center}
\end{figure}

To see why the code fragment is correct, consider that all the characters that precede occurrences of a valid phrase suffix $\alpha$ in the dataset are consecutive in the BWT.  If $\alpha$ is always preceded by the same distinct character then the code fragment just needs to fill in the right number of copies of that character; otherwise, it should leave enough blank cells to hold them.

The code fragment scans through the valid phrase suffixes in lexicographic order, by scanning through the suffixes of the dictionary and checking if each starts with a valid phrase suffix.  For each phrase $\beta$ in $\dict$ ending with $\alpha$,
\begin{itemize}
\item if $\alpha$ is always preceded by the same distinct character, then the code fragment fills into the BWT all the characters that precede $\alpha$ in occurrences of $\beta$ in the dataset;
\item otherwise, it leaves enough blank cells to hold all the characters that precede $\alpha$ in occurrences of $\beta$ in the dataset.
\end{itemize}
Since the number of characters that precede occurrences of $\alpha$ in the dataset is the sum over the phrases $\beta$ ending with $\alpha$ of the number of characters that precede $\alpha$ in occurrences of $\beta$ in the dataset, the code fragment is correct.

\section{Merging}
\label{sec:merging}

Suppose we have a dataset that can be broken down into small datasets that are not very similar to each other, in the sense that most $w$-mers that appear in the whole dataset appear in only one of the small datasets.  (In our experiments, we found setting $w = 20$ is usually enough.)  We find the set of trigger strings that appear in more than one small dataset, then perform a modified parsing of the small datasets treating as trigger strings only the ones appearing in exactly one small dataset.  Once we have the small datasets' BWTs we will be interested only in their dictionaries, so we can discard the parses.

Each valid phrase suffix now appears in the dictionary of exactly one of the small datasets.  The characters in the  whole dataset's BWT appear in the lexicographic order of the valid phrases suffixes they immediately precede in the whole dataset.  If two characters precede the same valid phrase suffix then they must be in the same small dataset and --- assuming the BWT of the whole dataset is an extended BWT (eBWT)~\cite{MRRS07} with the small datasets treated as separate strings or collections of strings --- their order in the whole dataset's BWT is the same as in that small dataset's BWT.

This means that if $\SAD$ is now the suffix array of the concatenation $\dict$ of the small dataset's dictionaries, $\dSize$, $\valid(\SAD[i])$ and $\occ(\SAD[i])$ are defined as before but $\BWT$ is now a file, and
\begin{itemize}
\item $\oldBWT[j]$ is a file containing the BWT of the $j$th small dataset,
\item $\DS(\SAD[i])$ is the number of the small dataset containing the phrase suffix starting at $\dict[\SAD[i]]$,
\item $\buffer$ is an empty string at least as long as the frequency of the most common valid phrase suffix,
\end{itemize}
then the code fragment in Figure~\ref{fig:merge_code} merges the small datasets' BWTs into the BWT of the whole dataset.

\begin{figure}[t]
\begin{center}
\begin{BVerbatim}
for (int i = 0; i < dSize; i++) {
    if (valid(SAD[i])) {
        fread(&buffer, 1, occ(SAD[i]), oldBWT[DS(SAD[i])]); 
        fwrite(&buffer, 1, occ(SAD[i]), BWT);
}   }
\end{BVerbatim}
\caption{A code fragment that merges the BWTs of the small datasets into the BWT of the whole dataset.}
\label{fig:merge_code}
\end{center}
\end{figure}

To see why the code fragment is correct, consider that for each valid phrase suffix $\alpha$ and phrase $\beta$ ending with $\alpha$, it copies as many characters from the BWT of the small dataset containing copies of $\alpha$ as there are occurrences of $\beta$ in that dataset.  The number of occurrences of $\alpha$ in the whole dataset is the total number of occurrences of phrases ending with $\alpha$, so the code fragment is correct.

The code fragment needs enough memory to store $\SAD$ and $\buffer$ and data structures to support $\valid(\SAD[i])$, $\occ(\SAD[i])$ and $\DS(\SAD[i])$.  The latter are easy to make small: we store
\begin{itemize}
\item a table saying which phrase in $\dict$ contains every, say, 50th character of $\dict$;
\item a table saying where each phrase ends in $\dict$;
\item a table saying how many times each phrase occurs in the whole dataset;
\item a table saying which dataset contains each phrase in $\dict$.
\end{itemize}

Given $\SAD[i]$, we use the first table to get a close lower bound on which phrase contains $\dict[\SAD[i]]$, then start at the corresponding number in the second table and scan until we find $\SAD[i]$'s successor to determine the exact phrase that contains $\dict[\SAD[i]]$.  We check that $\dict[\SAD[i]]$ is not the first character in the phrase (if so the phrase suffix starting at $\dict[\SAD[i]]$ is not proper) nor among the last $w - 1$ characters, to determine $\valid[\SAD[i]]$.  We look up $\occ[\SAD[i]]$ in the third table and $\DS[\SAD[i]]$ in the fourth table.

In practice, storing $\SAD$ takes less memory than storing the parse --- so our merging approach is uses less memory than building the BWT for the whole dataset directly with PFP --- but it still takes a lot.  We can simulate scanning $\SAD$ by building the suffix arrays of the small dataset's dictionaries and streaming them from disk, merging them to obtain $\SAD$ (and discarding each entry $\SAD[i]$ when we have finished the corresponding pass through the code fragment's loop).  To merge the suffix arrays, we can store the small dataset's dictionaries and perform {\tt strcmp()} queries to decide which suffix array points to the lexicographically next phrase suffix.  Storing the dictionaries instead of $\SAD$ reduces the space by about 4 or 8, depending on how large the suffix-array entries are.

If the BWT of the whole dataset will be used to support searches for queries over only $\{\mathtt{A}, \mathtt{C}, \mathtt{G}, \mathtt{T}\}$ then we can do somewhat better.  We replace all the non-$\{\mathtt{A}, \mathtt{C}, \mathtt{G}, \mathtt{T}\}$ characters in the small datasets by copies of {\tt X}, which does not change the multiset of substrings over $\{\mathtt{A}, \mathtt{C}, \mathtt{G}, \mathtt{T}\}$ in the whole dataset.  PFP uses {\tt 0x0} as an end-of-file symbol, {\tt 0x1} to mark the ends of phrases in its dictionaries and {\tt 0x2} to separate strings, but we can still store the small datasets' dictionaries in 3 bits per character.  Because phrase suffixes' encodings may start offset by different amounts into bytes, we can no longer use {\tt strcmp()}; with careful casting and bit-shifting we can compare 20-character blocks with one operation, however, so comparing phrase suffixes is still fairly fast.

Currently we keep a list of the small datasets' IDs, sorted into the lexicographic order of the phrase suffixes pointed to by those datasets' next suffix-array entries, and after we process a suffix-array entry we bubble its dataset's ID down the list to its new position.  When dealing with many small datasets, however, a min-heap should be more efficient.

We may be able to parallelize our merging algorithm efficiently, although we have not tried this yet.  We can choose a position partway through one of the small dataset's suffix array, say, then use binary search to find positions in the other small datasets' suffix arrays such that at some time in its execution, our algorithm be at all those positions in the suffix arrays simultaneously.  We can sum the previous $\occ(\SAD[i])$ values for each suffix array to find where our algorithm would be in the small datasets' BWTs at that time.  Therefore, we can start our algorithm from that time in its execution, and it follows we can parallelize our merging algorithm.  Notice the dictionaries and tables are static and can be shared between threads, so parallelization should not significantly increase our memory footprint.

\section{Experiments}
\label{sec:experiments}

We downloaded genomes of 30 species of bacteria from the AllTheBacteria project{\footnote{\url{https://allthebacteria.org/}}},
with sizes ranging from 309.93 MB to 314.53 MB and totalling 9.39 GB. We built their BWTs or eBWTs with our {\tt pfp-merge}\footnote{\url{https://gitlab.com/manzai/pfp-merge}}, Manzini's {\tt bigbwt}\footnote{\url{https://gitlab.com/manzai/Big-BWT}} (based on plain PFP), D\'iaz's {\tt grlbwt}\footnote{\url{https://github.com/ddiazdom/grlBWT}}, Li's {\tt ropebwt3}\footnote{\url{https://github.com/lh3/ropebwt3}} and Olbrich's {\tt lg}\footnote{\url{https://gitlab.com/qwerzuiop/lyndongrammar}} (for ``Lyndon grammar'').  We omitted Masillo's CMT-BWT because it requires a single reference. We used default parameter settings for all the programs (except we used {\tt ropebwt3}'s {\tt -R} flag to prevent it indexing also the datasets' reverse complements) and, due to time constraints, 16 threads for each program on similar nodes of a cluster at the University of Helsinki.  For {\tt pfp-merge}, {\tt bigbwt}, and {\tt lg}, we removed newlines and converted non-$\{\mathtt{A}, \mathtt{C}, \mathtt{G}, \mathtt{T}\}$ characters to {\tt X}s. We ran {\tt ropebwt3} on the original FASTA files and {\tt grlbwt} on the concatenation of the datasets (one string per line).  Table~\ref{tab:results} shows the wall-clock time and peak memory used by each program. Although {\tt pfp-merge} was almost the slowest --- which may improve with additional parallelization --- it used the least memory by a noticeable margin.  The actual merging used only 0.77 GB of memory, suggesting we may perform relatively even better on more datasets.

\begin{table}[t!]
\begin{center}
\caption{The time and peak memory used by each program on 30 bacterial pangenomes totalling 9.39 GB.}
\label{tab:results}
\begin{tabular}{r|r|r}
& \multicolumn{1}{c|}{time} & \multicolumn{1}{c}{memory} \\
& \multicolumn{1}{c|}{(mm:ss)} & \multicolumn{1}{c}{(GB)} \\
\hline & & \\[-1.5ex]
{\tt pfp-merge} & 23:50 & 0.91 \\[.5ex]
{\tt bigbwt} & 16:26 & 13.84 \\[.5ex]
{\tt glrbwt} & 9:01 & 1.71 \\[.5ex]
{\tt ropebwt3} & 27:10 & 5.12 \\[.5ex]
{\tt lg} & 7:41 & 7.00
\end{tabular}
\end{center}
\end{table}

\begin{credits}
\subsubsection{\ackname}

Many thanks to Lavinia Egidi, Felipe Louza and Giovanna Rosone for helpful discussions, and belated thanks to Pawe\l\ Gawrychowski, Dmitry Kosolobov and Tatiana Starikovskaya for suggestions that sped up PFP in the past.  TG funded by NSERC grant RGPIN-07185-2020.  BL and VS funded by NIH grant R01HG011392 to BL.  ZsL partially funded by MUR PRIN project no.\ 2022YRB97K and INdAM-GNCS project no.\ E53C24001950001. VG and GM funded by the Next Generation EU PNRR MUR M4 C2 Inv 1.5 project ECS00000017 Tuscany Health Ecosystem Spoke 6 CUP I53C22000780001.

\subsubsection{\discintname}

The authors declare no competing interests.

\end{credits}

\end{document}